# Detecting Atomic Scale Surface Defects in STM of TMDs with Ensemble Deep Learning

Darian Smalley[1,2,*], Stephanie D. Lough[1,2], Luke Holtzman[3], Kaikui Xu[4], Madisen Holbrook[3], Matthew R. Rosenberger[4], J.C. Hone[3], Katayun Barmak[3], Masahiro Ishigami[1,2]

[1]Department of Physics, UCF, 4111 Libra Drive, PS430 Orlando, FL 32816, USA

[2]NanoScience Technology Center, UCF, 12424 Research Parkway, Suite 400, Orlando, FL 32826, USA

[3]Department of Applied Physics and Applied Mathematics, University of Columbia, 500 W. 120th Street, Suite 200, New York, NY 10027, USA

[4]Department of Aerospace and Mechanical Engineering, University of Notre Dame, 257 Fitzpatrick Hall of Engineering, Notre Dame, IN 46556, USA

*Phone: (407) 989-2087, Fax: (407) 823-5112, Email: darian.smalley@ucf.edu

## Abstract

Atomic-scale defect detection is shown in scanning tunneling microscopy images of single crystal $WSe_2$ using an ensemble of U-Net-like convolutional neural networks. Standard deep learning test metrics indicated good detection performance with an average $F_1$ score of 0.66 and demonstrated ensemble generalization to C-AFM images of $WSe_2$ and STM images of $MoSe_2$. Defect coordinates were automatically extracted from defect detections maps showing that STM image analysis enhanced by machine learning can be used to dramatically increase sample characterization throughput.

**Introduction**

The relevance of atomic-scale defects on the physical properties of semiconductor devices is well known [1]. This is especially true for devices fabricated from two-dimensional (2D) transition metal dichalcogenides (TMDs), such as tungsten diselenide ($WSe_2$) [2]. Point defect detection in 2D TMDs, particularly at the device size scale, is still in its infancy due to the challenges these materials pose to commonly used techniques such as electron microscopy [3] and optical microscopy [4]–[6]. Scanning probe microscopy (SPM) techniques, like scanning tunneling microscopy (STM) and conductive atomic force microscopy (C-AFM), can directly image atomic-scale defects in TMD devices but the routine manual image analysis required is a major experimental throughput bottleneck.

Deep learning (DL) has been shown to vastly enhance the capabilities of scanning probe microscopy (SPM) [7]–[12]. The analysis of STM images using DL is challenging due to the inherent noise and instability arising from the tip condition and tunneling current contrast mechanism. We report results on applying deep learning to the automatic detection of atom-scale defects in STM images of single crystal $WSe_2$. Our results show that an ensemble of U-Net models [13], a convolutional neural network (CNN) which has seen a widespread success of its application to many computer vision problem domains, can overcome challenges imposed by manual defect counting.

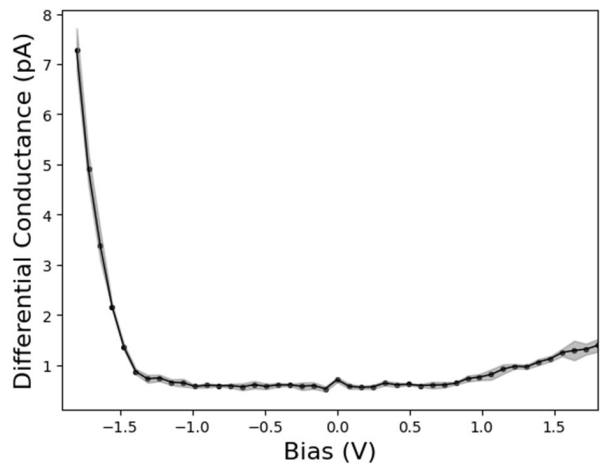

**Fig. 1** Average differential conductance of pristine $WSe_2$ with standard deviation shown as the shaded regions showing n-type character. STM training images were aquired at -1.4 V

**Materials & Methods**

CVT and Flux Growth of WSe$_2$ Crystals

All WSe$_2$ crystals used for this study were synthesized by the self-flux method following previously published methods [14].

Scanning Tunneling Microscopy

The STM images of single crystal WSe$_2$ discussed and used in this work were acquired using -1.4 V bias voltage and 60 pA tunneling current by a ScientaOmicron Low Temperature STM (LT-STM) with Specs Zurich GmbH's 'Nanonis' SPM controls. This bias was chosen to maximize the sensitivity of STM to charged defects due to its location near the valence band edge, as seen in Figure 1, and we empirically observe the most types of defects at this voltage. The sample temperature and pressure during imaging were approximately 77 K and $7.2 \times 10^{-12}$ Torr. STM images were acquired at three scan widths of 50 nm, 100 nm, and 200 nm and three resolutions of 512x512, 1024x1024, and 2048x2048. We used images with nm-to-pixel ratios of 0.049 and 0.098. Electrochemically etched W tips were used as the STM probe. These probes were processed in situ by electron-beam heating followed by conditioning on epitaxial Au(1 1 1) on mica. Conditioning concluded once the well-known Au(1 1 1) herringbone reconstruction and the corresponding surface state in scanning tunneling spectra were resolved [15]. Bulk WSe$_2$ crystals were attached to the STM sample plates using Epo-Tek H20E silver conductive epoxy along with a ceramic post affixed to the top surface of the crystal. The crystal was exfoliated in-situ without exposure to oxygen using the ceramic post in an adjacent preparation chamber to expose a fresh, clean crystal surface. The tip was aligned to the sample optically.

Data Preprocessing and Augmentation

The training data included both atomic resolution as well as non-atomic resolution STM images which were plane corrected, quadratically flattened line-by-line, gaussian smoothed, and equalized with contrast limited adaptive histogram equalization to improve feature clarity. These operations were performed using the NumPy and Open-Source Computer Vision packages in Python 3. Data augmentation was accomplished by randomly cropping 60 256 * 256 resolution regions from 38 labeled, parent 512 * 512 resolution, 50 nm * 50 nm STM images covering 0.1 um2 total area. No linear transformations were applied to avoid introducing non-physical artifacts that can reduce the model's applicability to STM images. These data were then split into three groups: the training set, the validation set, and the test set. The training and validation sets were used to train the U-Net ensemble and optimize hyperparameters, while the test set was used post-training to measure how well the ensemble learned to generalize.

Deep Learning

Deep learning ensemble creation and training was accomplished using AtomAI, the open-source Pytorch-based package for deep/machine learning analysis of microscopy data using Python 3 [12]. An ensemble of models was trained instead of a single model to obtain information on the variance in detection confidence amongst the 10 U-Net models. Cross entropy loss and the ADAM optimizer were used for training. The following training parameters were used after several rounds of hyperparameter optimization: 500 epochs, learning rate $10^{-3}$, and batch size 1. After 500 epochs, the weights of each model were stochastically averaged. Training typically took 40 minutes on a single desktop with an Intel i9-11900, Nvidia RTX 1070, and 32 GB of 3200 MHz DDR4 RAM.

Training Data and Ensemble Training

Figure 2a shows a typical input image used for training the neural networks. Individual defect instances were manually annotated with bounding boxes and associated qualitative labels to form a dataset of 1382 instances of $WSe_2$ defects. We classified defects as either peaks or troughs in intensity to train our ensemble. The bounding boxes were used to generate segmentation masks for each STM image where a circle was placed at the center of each bounding box with a characteristic width approximated from manual inspection of height profiles of a small subset of defects. Every pixel within each circle was set to 1 for the trough class or 2 for the peak class. Pixels which correspond to the atomic or noise background were set to 0. An example segmentation mask generated from Figure 2a is shown in Figure 2b, where the blue color signifies trough-like defects while the yellow color represents peak-like defects. Then, the experimental images and masks were used to train the ensemble of 10 U-Net models. Once training was completed, the U-Net ensemble was applied to full STM images to predict defect segmentation maps, as shown in Figure 2c. Detection variance was greatest around the perimeter of charged defects due to the gradual decrease in contrast resulting from the charge halo surrounding these defects. Figure 2d shows the coordinates extracted from the defect segmentation map shown in Figure 2c overlayed on the STM image shown in Figure 2a.

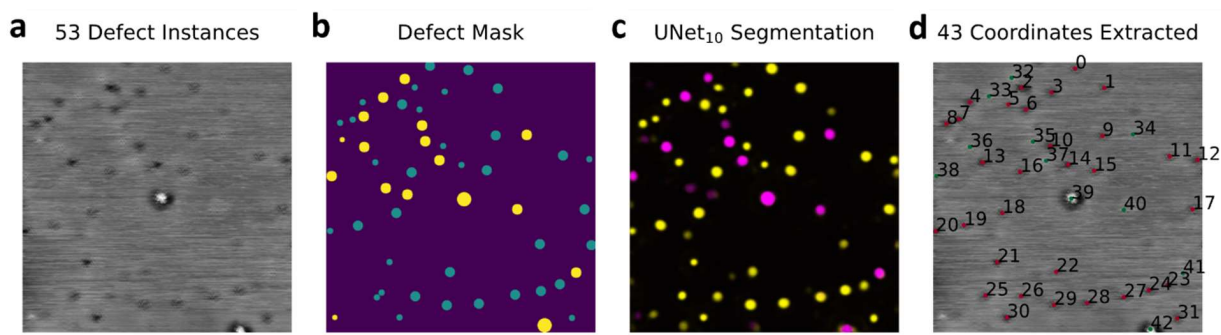

**Fig. 2** Example of an input, output, and coordinate extraction from ensemble. **(a)** Example 50 nm * 50 nm STM image (-1.4 V, 60 pA). **(b)** Segmentation mask produced from the bounding boxes.

Blue is the trough class while yellow is the peak class. These masks were used to train the U-Net model. **(c)** U-Net segmentation map predicted after training was completed. The yellow color signifies trough-like defects while the purple color represents peak-like defects. **(d)** Defect XY coordinates extracted from the U-Net segmentation map

**Results & Discussion**

The generalization performance of the trained ensemble was evaluated using 100 un-seen, labeled test images of $WSe_2$ to compute the confusion matrix, the receiver-operator-characteristic curve, the precision-recall curve, mean average precision, and the $F_1$-score as shown in Figure 3. Figure 3a shows a normalized confusion matrix computed over the test images, without inclusion of the background class, which displays an almost perfect classifier with high values along the diagonal and low values on the off diagonal. This matrix shows that the trained ensemble does not display classification bias despite vast class imbalance in the training data dominated by trough defects. The precision-recall curve is shown in Figure 3b, where a perfect classifier would maximize both precision and recall and achieve a high $F_1$ score. Here the mean average precision (mAP) of the peak and trough classes were above the 0.5 baseline required of a trustworthy classifier and achieved $F_1$ scores of 0.69 and 0.63, respectively. The receiver-operator characteristic (ROC) curve shown in Figure 3c displays a high true positive rate at a low false positive rate for every class with an area under the ROC (AUROC) close to 1, far higher than random guessing which would result in an AUROC of 0.5.

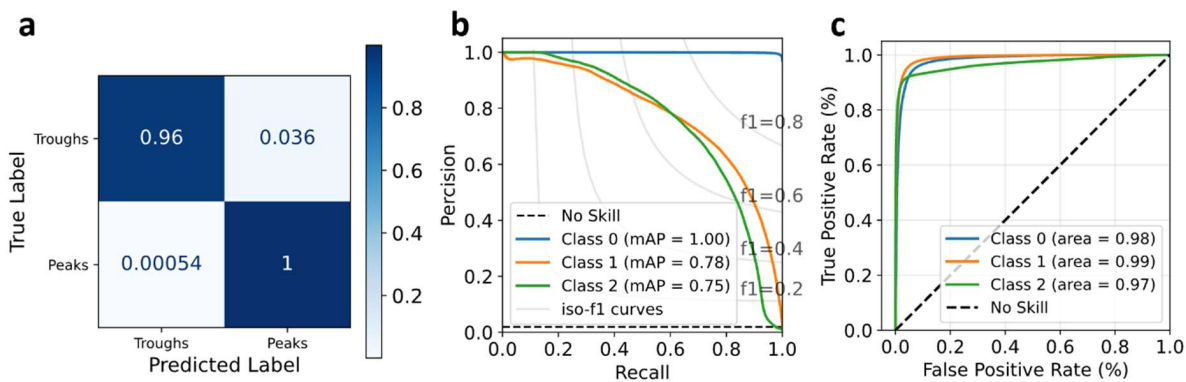

**Fig. 3** U-Net Ensemble Evaluation on 100 Unseen Test STM Images **(a)** Normalized confusion matrix displaying good generalization of peak and trough detection indicated by high values along the diagonal. **(b)** Precision-Recall Curves of each class showing a class averaged mAP = 0.78 and $F_1$ = 0.66. **(c)** Receiver operator characteristic curves showing a class averaged AUROC = 0.97. Class 0, 1, and 2 are background, troughs, and peaks, respectively

While the U-Net ensemble was trained on STM images of $WSe_2$, we found that it demonstrated generalization to images of $WSe_2$ acquired using C-AFM, and STM images of $MoSe_2$, as shown in of single crystal MoSe$_2$ aquired at -1.9 V **d)** Defect coordinates extracted from the defect detection map produced using the STM image shown in Figure 4c as input to the trained ensemble.

Figure 4. Figure 4a shows a 99 nm * 99 nm C-AFM image of single crystal WSe$_2$ from a similar crystal batch with 1024 * 1024 pixel resolution. Figure 4b shows the coordinates extracted from the defect detections produced by the U-Net ensemble overlayed on the input image shown in Figure 4a. The red dots correspond with peak-type defects while the green dots represent the trough-type defects. While the detections are not perfect, the ensemble displays remarkable generalization to C-AFM images without any re-training on new labeled data. We attribute this capability to the similarity of the image contrast mechanism between C-AFM and STM [16]. Figure 4c shows a 50 nm * 50 nm STM image of MoSe$_2$ aquired at -1.9 V with 512 * 512 resolution. Similar to Figure 4b, Figure 4d shows the coordinates of defects detected by the U-Net ensemble overlayed on the input STM image shown in Figure 4c. Here the generalization of the trained ensemble to STM images of MoSe$_2$ is attributed to the similarity of point defect appearances across 2H TMD material systems which include selenium.

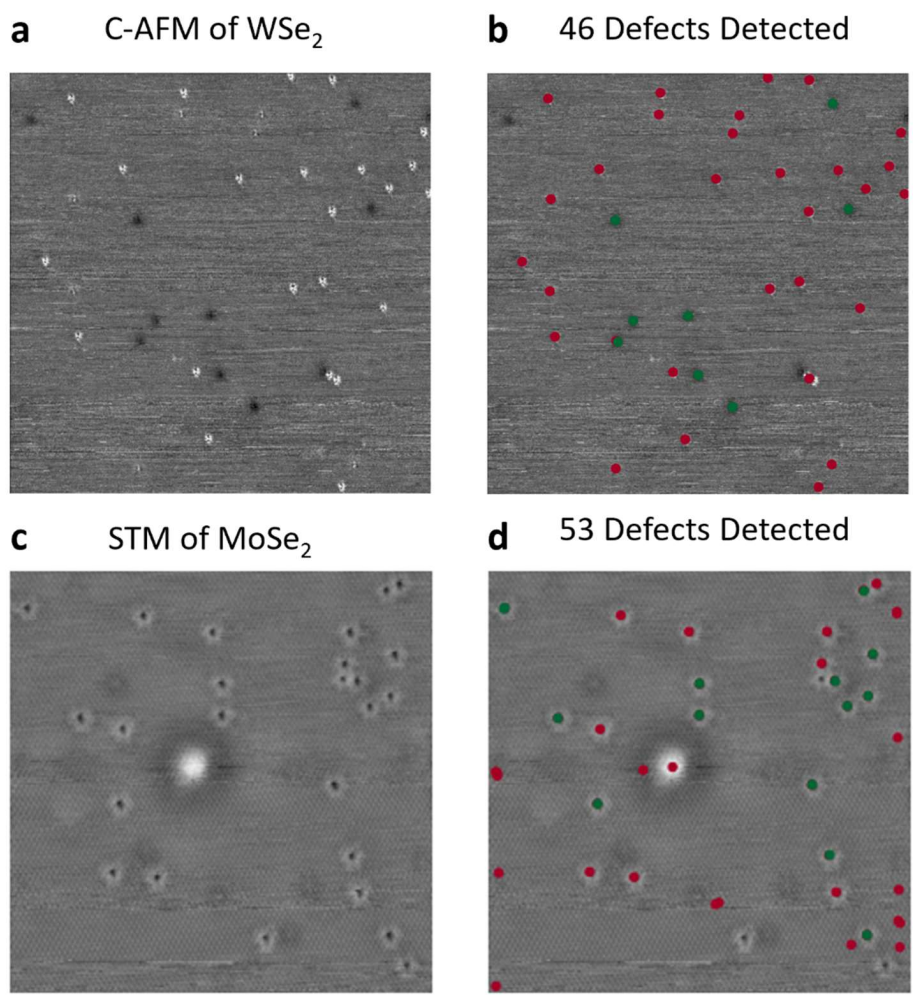

**Fig. 4** Example U-Net Ensemble Generalization to C-AFM and MoSe$_2$ **a)** 1024 * 1024 resolution, 99 nm x 99 nm C-AFM image of single crystal WSe$_2$. **b)** Detected defect coordinates overlayed on top of the C-AFM image shown in figure 4a. **c)** 512 * 512 resolution, 50 nm * 50 nm STM image

## Conclusion

We have applied deep learning to automatically detect atomic-scale defects in STM images of single crystal WSe$_2$. We found that an ensemble of models with the U-Net architecture could reliably detect defects in STM images of this crystal and could generalize to other SPM methods and other materials. This represents another step towards the data-driven optimization of TMD synthesis and characterization.

## Authors Contributions

All authors contributed to the conception, design, and implementation of the study. Material synthesis was performed by Luke Holtzman. Material preparation was performed by Stephanie D. Lough. STM data collection of WSe$_2$ was performed by Darian Smalley. C-AFM data collection was performed by Kaikui Xu. STM data collection of MoSe$_2$ was performed by Madisen Holbrook. All deep learning and data analysis was performed by Darain Smalley. The first draft of the manuscript was written by Darian Smalley. All authors commented on previous versions of the manuscript. All authors read and approved the final manuscript.

## Conflict of Interest

On behalf of all authors, the corresponding author states that there is no conflict of interest.

## Data Availability

Data is available from the authors upon reasonable request.

## Funding

Not applicable.